\documentclass{article}
\usepackage{amsfonts,amssymb, amsmath}
\usepackage[english]{babel}

\DeclareMathAlphabet{\mathpzc}{OT1}{pzc}{m}{it}

\def\nn{\nonumber }
\def\bq{ \begin{equation} }
\def\eq{ \end{equation} }
\def\ben{ \begin{eqnarray} }
\def\en{ \end{eqnarray} }

\def\e{{\rm e}}

\newtheorem{prop}{Proposition}

\begin{document}


\title{On bi-hamiltonian structure of some integrable systems on  $so^*(4)$.}
\author{A V Tsiganov \\
\it\small
St.Petersburg State University, St.Petersburg, Russia\\
\it\small e--mail: tsiganov@mph.phys.spbu.ru}

\date{}
\maketitle

\begin{abstract}
We classify  quadratic Poisson structures on $so^*(4)$ and $e^*(3)$, which have the same foliations by symplectic leaves as  canonical Lie-Poisson tensors. The separated variables for  some of the corresponding bi-integrable systems are constructed.
\end{abstract}

\par\noindent
PACS: 45.10.Na, 45.40.Cc
\par\noindent
MSC: 7020; 70H06; 37K10

\vskip0.1truecm

\section{Introduction}
\setcounter{equation}{0}
Let $M$ be a Poisson manifold endowed with a  bivector $P_0$ fulfilling the Jacobi condition
\[
[P_0,P_0]=0
\]
with respect to the Schouten bracket $[.,.]$ \cite{vais93}.

A bi-Hamiltonian manifold $M$ is a smooth  (or complex) manifold endowed with two compatible bivectors $P_0,P_1$ such that
\bq\label{m-eq}
[P_0,P_0]=[P_0,P_1]=[P_1,P_1]=0.
\eq
Classification of compatible Poisson bivectors and the corresponding bi-in\-teg\-rab\-le systems with integrals of motion
\[
\{H_i,H_j\}_0=\{H_i,H_j\}_1=0,\qquad  i,j=1,\ldots,n,
\]
is nowadays a subject of  intense research \cite{fp02,ts07b,ts07c}. Of course, direct solution of the equations (\ref{m-eq}) is generally quite difficult. We can try to lighten this work by using properties of the given Poisson manifold $(M,P_0)$.

Bivectors $P_1$ fulfilling the compatibility condition $[P_0,P_1]=0$ are called 2-cocycles in the Poisson-Lichnerowicz cohomology defined by $P_0$ on $M$ \cite{lih77}.  The Lie derivative of $P_0$ along any vector field $X$ on $M$
\bq\label{co-b}
P_1=\mathcal L_X(P_0)
\eq
is 2-coboundary, more precisely it is 2-cocycle associated with the Liouville vector field  $X$. For such bivectors   $P_1$ the system of equations (\ref{m-eq}) is reduced to the single  equation
\bq\label{m-eq2}
[\mathcal L_X(P_0),\mathcal L_X(P_0)]=0,\quad\Leftrightarrow\quad\
[\mathcal L_X^2(P_0),P_0]=0.
\eq

The second Poisson-Lichnerowicz cohomology group $H^2_{P_0}(M)$ of $M$ is precisely the set of bivectors $P_1$ solving $[P_0,P_1]=0$ modulo the solutions of the form $P_1=\mathcal L_X(P_0)$. We can interpret  $H^2_{P_0}(M)$ as the space of infinitesimal deformations of the Poisson structure modulo trivial deformations.  For regular Poisson manifolds this cohomology reflects the topology of the leaf space and the variation in the symplectic structure as one passes from one leaf to another  \cite{vais93}.

The aim of this paper is to show some particular solutions of the equation (\ref{m-eq2})
and to discuss separation of variables  for the corresponding bi-integrable system on  $so^*(4)$.

\section{The Lie-Poisson bivectors on $so^*(4)$.}
\setcounter{equation}{0}

Let us consider the semisimple Lie algebra $so(4)$. The dual space
$M=so^*(4)$ is  the  Poisson manifold \cite{vais93}.
Since $M=so^*(4)=so^*(3)\oplus so^*(3)$, we can introduce the following coordinates $z=(s,t)$ on $M$, where  $s=(s_1,s_2,s_3)$ and $t=(t_1,t_2,t_3)$
are two vectors in $\mathbb R^3$. As usual we identify
$(\,\mathbb R^3,\wedge)$ and  $(so(3),[.,.])$ by using the well known isomorphism of the Lie algebras
\bq\label{trans-M}
 z=\left(z_1,z_2,z_3\right)\to z_M=\left(\begin{smallmatrix}
            0 & z_3 & -z_2 \\
            -z_3 & 0 & z_1 \\
            z_2 & -z_1 & 0
          \end{smallmatrix}\right),
\eq
where $\wedge$ is the cross product in $\mathbf R^3$ and  $[.,.]$ is the matrix commutator in $so(3)$. In these coordinates the canonical Poisson bivector on  $so^*(4)$
is equal to
\bq \label{p0}
 P_0=\left(\begin{array}{cc}s_M&0\\0&
t_M\end{array}\right).
\eq
The generic symplectic leaves are  the level sets of two globally defined Casimir functions
\bq\label{Caz}
C_1=\langle s,s\rangle\equiv|s|^2\equiv\sum_{i=1}^3s_i^2,\qquad C_2=\langle t,t \rangle, \qquad P_0dC_{1,2}=0.
\eq

In the paper \cite{ts06} the overdetermined system of equations (\ref{m-eq}) on $so^*(4)$ has been directly solved in the class of the linear Lie-Poisson  bivectors $P_1$, i.e using the nzats $P_1^{ij}=\sum a^{ij}_m z_m$ for the components of the second bivector.

According to \cite{ts06} the  Frahm-Schottky, Steklov and Poincar\'{e} systems are bi-integrable systems  and the corresponding second bivectors $P_1=\mathcal L_X(P_0)$ are generated by the  Liouville vector fields   $X_j=\sum X^i_j\partial_i$ with the following components
\[
\begin{array}{l}
X_1=(a_1t_1,a_2t_2,a_3t_3,a_1s_1,a_2s_2,a_3s_3),\\
\\
X_2= (\mathcal X_1,a \mathcal X_2)\\
\\
X_3=(0, 0, at_3, 0, 0, -as_3)\,,
\end{array}\qquad a,a_1,a_2,a_3\in\mathbb C,
\]
where
\[
\begin{array}{l}
\mathcal X_1=(
a_1t_1+\frac{a_1(a_2^2+a_3^2)}{a_2a_3}s_1,
a_2t_2+\frac{a_2(a_1^2+a_3^2)}{a_1a_3}s_2,
a_3t_3+\frac{a_3(a_1^2+a_2^2)}{a_1a_2}s_3),\\
\\
\mathcal X_2=(-a_2a_3s_1-\frac{a_2^2+a_3^2}{2}t_1,-a_1a_3s_2-\frac{a_1^2+a_3^2}{2}t_2,
-a_1a_2s_3-\frac{a_1^2+a_2^2}{2}t_3),
\end{array}
\]
The remaining four solutions of the system (\ref{m-eq}) are associated with the following vector fields
\[\begin{array}{l}
X_4=((a_3+a_2)s_1, (a_3+a_1)s_2, (a_1+a_2)s_3, (b_2+b_3)t_1, (b_1+b_3)t_2, (b_1+b_2)t_3)\\
\\
X_5=(-\frac{a}2s_1+b_1t_1, -\frac{a}2s_2+b_2t_2, b_3t_3, b_1s_1-\frac{a}2t_1, b_2s_2-\frac{a}2 t_2, (b_3-a)s_3-at_3)\\
\\
X_6=(a_2s_1, a_1s_2, (a_1+a_2)s_3-bt_3, ct_1, ct_2, 0)\\
\\
X_7=(as_1, as_2, 0, c_1s_3+bt_1, c_2s_3+bt_2, 0),
\end{array}
\]
where $b_1=\pm b_2$ for the vector field $X_5$. The corresponding bi-integrable systems with quadratic integrals of motion have been completely described in \cite{ts06}.

We have to keep in the mind that
the vector fields $X_k$ and the corresponding bivectors $P_1=\mathcal L_{X_k} P_0$ are defined up to  canonical transformations  of the vectors $s$ and $t$, which preserve the form of the canonical bivector $P_0$ (\ref{p0}) \cite{kst03}.

\section{Quadratic Poisson bivectors  on $so^*(4)$ and $e^*(3)$}
\setcounter{equation}{0}
After the linear Poisson structures, it is natural to look at quadratic structures.
Substituting the following anzats for the components of the Liouville vector fields $X_m=\sum_{i\geq j}^6 a_m^{ij} z_iz_j$ into (\ref{m-eq2}) one gets
a highly overdetermined system of quadratic equations on the 126 coefficients $a_m^{ij}$. Unfortunately, we cannot get and classify
all the solutions of this system even by using modern computers and modern software.

So, in this section we will suppose that
\bq\label{m-eq3}
P_1dC_{1,2}=0.
\eq
It means that symplectic leaf of $P_1$ are contained in those of $P_0$. In general
bivector $P_1$ could have some more Casimirs, so that their symplectic leaf could be smaller. For example,  if $X=0$, then equation (\ref{m-eq3}) is true, but the symplectic foliations of $P_0$ and $P_1$ are different. However in this case $P_1=0$ does not linear or quadratic bivector on $so^*(4)$.

\begin{prop}
For all the quadratic bivectors $P_1=\mathcal L_X(P_0)$ on $so^*(4)$  restriction (\ref{m-eq3}) leads to quadratic Poisson structures having the same foliation by symplectic leaves as $P_0$ only.
\end{prop}
We will prove  this Proposition by using so-called Darboux-Nijenhuis variables.

The additional restriction on $P_1=\mathcal L_X(P_0)$ is a linear equation with respect to  $X$, which allows us to get solutions of the following system of equations
\bq\label{sys-meq}
\mathcal L_X(P_0)dC_{1,2}=-P_0dX(C_{1,2})=0\qquad\mbox{\rm and}\qquad
[\mathcal L_X(P_0),\mathcal L_X(P_0)]=0
\eq
Any solution $X$ of this system (\ref{sys-meq}) gives rise to a Poisson bivector $P_1=\mathcal L_X(P_0)$ fulfilling  equations (\ref{m-eq}) and (\ref{m-eq3}).

We solved equations (\ref{sys-meq}) in the class of quadratic vector fields $X=\sum X_m\partial_m$ with the components $X_m=\sum_{i\geq j} a_m^{ij} z_iz_j$ by using one of the modern computer algebra system and got  the following three solutions.

\begin{prop}
Let $a$ and $b$ are numerical vectors and $\alpha,\beta$ are
arbitrary parameters. The vector field  $X=(\mathcal X_1,\mathcal X_2)$ with  components
\[
\mathcal X_1=\alpha\bigl(s\wedge(a\wedge s)\bigr)+\beta\langle b,t\rangle(s\wedge a),
\qquad \mathcal X_2=\alpha\bigl(t\wedge(b\wedge t)\bigr)\,
\]
generates  a first bivector fulfilling  equations
(\ref{m-eq}) and (\ref{m-eq3})
\bq\label{p1-1}
P_1^{(1)}=\left(
                        \begin{matrix}
                          2\alpha\langle a,s\rangle s_M & \beta\bigl[(a\wedge s)\otimes(b\wedge t)\bigr]  \\
                         -\beta\bigl[(a\wedge s)\otimes(b\wedge t)\bigr]^T &
                         2\alpha\langle b,t\rangle t_M
                        \end{matrix}
                      \right).
\eq
\end{prop}
Using orthogonal transformations of the vectors
\bq\label{ort-tr}
s\to
s'=U_1s,\qquad\mbox{and}\qquad t\to t'=U_2t
\eq
where $U_{1,2}$ are orthogonal  matrices, we can always put
 \[a=(0,0,a_3),\qquad b=(0,0,b_3).\]
Using scaling transformation
\bq\label{s-tr}
P_1^{(1)}\to \lambda P_1^{(1)},\qquad \lambda\in \mathbb R\,,
\eq
we can put
$\alpha=1$ or $\beta=1$.
\begin{prop}
Let $a$ and $b$ are two complex vectors, such that
\[
 \langle a,b\rangle=\langle b,b\rangle=0,
\]
where $\langle x,y\rangle=\sum_{i=1}^3 x_iy_i$ is the inner product of two vectors.
The vector field  $X=(\mathcal X_1,\mathcal X_2)$ with  components
\[
\begin{array}{l}
\mathcal X_1=\frac{1}{2}\,s\wedge(a\wedge s),\\
\\
\mathcal X_2=-\frac{1}{2}\,t\wedge(a\wedge t)
+ i|a|^{-1}\langle a,s\rangle(a\wedge t)
-i\langle b,s\rangle(b\wedge t),\end{array}
\]
generates a second bivector  fulfilling  equations (\ref{m-eq}) and
(\ref{m-eq3})
\bq\label{p1-2}
P_1^{(2)}=\left(
                        \begin{smallmatrix}
                          \langle a,s\rangle s_M & -i\bigl[|a|^{-1}(a\wedge s)\otimes(a\wedge t)-(b\wedge s)\otimes(b\wedge t)\bigr]  \\
                         i\bigl[|a|^{-1}(a\wedge s)\otimes(a\wedge t)-(b\wedge s)\otimes(b\wedge t)\bigr]^T &
                         -\langle a,t\rangle t_M
                        \end{smallmatrix}
                      \right),
\eq
here $|a|=\sqrt{a_1^2+a_2^2+a_3^2}$ and $(x\otimes y)_{ij}=x_iy_j$.
\end{prop}
Using orthogonal transformations (\ref{ort-tr}) and scaling
(\ref{s-tr})
 we can always put
 \[
a=(0,0,1),\qquad b=(b_1,ib_1,0)\,.
 \]
\begin{prop}
Let $a,b$ and $c$ are three complex vectors, such that
\[
 \langle a,b\rangle=\langle a,c \rangle=\langle b,b\rangle=\langle c,c\rangle=0,\qquad
b\wedge c\neq0.
\]
The vector field  $X=(\mathcal X_1,\mathcal X_2)$ with  components
\[
\begin{array}{l}
\mathcal X_1=\frac{1}{2}\,s\wedge(a\wedge s),\\
\\
\mathcal X_2=\frac{1}{2}\,t\wedge(a\wedge t)
-i|a|^{-1}\langle a,s\rangle(a\wedge t)
-i\langle b,s\rangle(c\wedge t)
,\end{array}
\]
generates a third bivector  fulfilling  equations (\ref{m-eq}) and
(\ref{m-eq3})
\bq\label{p1-3}
P_1^{(3)}=\left(
                        \begin{smallmatrix}
                          \langle a,s\rangle s_M & i\bigl[|a|^{-1}(a\wedge s)\otimes(a\wedge t)+(b\wedge s)\otimes(c\wedge t)\bigr]  \\
                         -i\bigl[|a|^{-1}(a\wedge s)\otimes(a\wedge t)+(b\wedge s)\otimes(c\wedge t)\bigr]^T &
                         \langle a,t\rangle t_M
                        \end{smallmatrix}
                      \right).
\eq
\end{prop}
Using orthogonal transformations (\ref{ort-tr}) and scaling
(\ref{s-tr})
 we can always put
 \[
a=(0,0,1),\qquad b=(b_1,ib_1,0),\qquad c=(c_1,-ic_1,0)\,.
 \]

Let us consider the canonical bivector on $M=e^*(3)$
 \bq \label{p0e}
 P_0=\left(\begin{array}{cc}0&x_M\\x_M&
J_M\end{array}\right)
\eq
and  the corresponding Casimir functions
\[
C_1=|x|^2=x_1^2+x_2^2+x_3^2,\qquad C_2=\langle x,J\rangle=x_1J_1+x_2J_2+x_3J_3.
\]
On $M=e^*(3)$ the system of equations (\ref{sys-meq})  has only one nontrivial solution in the class of quadratic vector fields $X=\sum X_m\partial_m$  with the components $X_m=\sum_{i\geq j} a_m^{ij} z_iz_j$, where $z=(x_1,x_2,x_3,J_1,J_2,J_3)$.
\begin{prop}
If $a$ and $b$ are two vectors, such that $|a|=0$,  and $\alpha$ is arbitrary parameter,
then the following bivector on $M=e^*(3)$
\bq\label{p1-e}
P_1^{(4)}=\left(\begin{smallmatrix}
\langle a,x\rangle,\qquad& (x\wedge J)\otimes a+\langle a,J\rangle x_M+\frac{1}{2}\left(\frac{1}{\alpha}+\alpha\langle a,b\rangle\right)\bigl(x\otimes x -|x|^2\bigr) +\alpha(b\wedge x)\otimes (a\wedge x)
\\
\\
\\
*,
& \langle a,J\rangle J_M+\langle b,x \rangle x_M
+\frac{1}{2}\left(\frac{1}{\alpha}+\alpha\langle a,b\rangle \right)(x\wedge J)_M
-\alpha\bigl((a\wedge x)\wedge(b\wedge J)\bigr)_M
\end{smallmatrix}\right)
\eq
satisfies equations (\ref{m-eq}) and (\ref{m-eq3}).
\end{prop}
This bivector $P_1^{(4)}$ is a Lie derivative of $P_0$. For the brevity we omit the explicit expression for the corresponding Liouville vector field $X$.

In the next Section we prove
that the Poisson bivectors $P_1^{(m)}$  have a common symplectic foliation with $P_0$
and that the Poisson bivectors $P_1^{(2)}$ and $P_1^{(3)}$ are equivalent.

\section{The Darboux-Nijenhuis variables}
\setcounter{equation}{0}

Let us consider a bi-hamiltonian manifold $M$
with the non degenerate  Poisson bivectors  $P_0$ and $P_1$.
By definition a set of local coordinates $(q_i, p_i)$ on $M$ is called
a set of Darboux-Nijenhuis coordinates if they
are canonical with respect to the symplectic form
\[\omega=P_0^{-1}=\sum_{i=1}^n dp_i\wedge dq_i
\]
and put the recursion operator $N=P_1P_0^{-1}$ in diagonal form,
\bq\label{xy-pedr}
 N=\sum_{i=1}^n q_i\left(
\dfrac{\partial }{\partial  q_i}\otimes dq_i+ \dfrac{\partial
}{\partial p_i}\otimes dp_i\right),.
\eq
This means that the only nonzero Poisson brackets are
\bq\label{br-xy}
\{q_i,p_j\}_0=\delta_{ij},\qquad \{q_i,p_j\}_1=q_i\delta_{ij},\qquad
\{q_i,q_j\}_{0,1}=\{p_i,p_j\}_{0,1}=0\,.
\eq
According to (\ref{xy-pedr}) coordinates $q_i$ are eigenvalues of $N$, i.e they are roots of the minimal  polynomial $\mathcal A(\lambda)=\bigl(\det(N-\lambda\mathrm I) \bigr)^{1/2}$ of $N$.

In order to get momenta $p_{1,2}$ we can directly solve equations (\ref{br-xy}) with respect to the functions $p_{1,2}(s,t)$. The standard computation problem is that variables  $p_{1,2}$  are
defined up to canonical transformation $p_i\to p_i+f_i(q_i)$, where
$f_i$ are arbitrary functions on $q_i$.

In order to construct recursion operator  $N$ on the generic symplectic leaves of $so^*(4)$ we will use analog of the Andoyer variables \cite{and23,bm}.

\subsection{Analog of the Andoyer variables on $so^*(4)$}
Let us introduce the following analog of the Andoyer variables \cite{and23}
\bq\label{u1}
u_1=s_3+t_3,\qquad
v_1=-i\ln\left(\dfrac{s_2+is_1+t_2+it_1}{\sqrt{(t_1+s_1)^2+(s_2+t_2)^2}}\right)
\eq
and
\ben
u_2&=&\sqrt{C_1+C_2+2s_1t_1+2s_2t_2+2s_3t_3},\nn\\
\label{u2}\\
v_2&=&\arccos\left(
\dfrac{t_3C_1-s_3C_2+(t_3-s_3)(s_1t_1+s_2t_2+s_3t_3)}
{\sqrt{(t_1+s_1)^2+(s_2+t_2)^2}\sqrt{C_1C_2-(s_1t_1+s_2t_2+s_3t_3)^2}}
\right)\nn
\en
The inverse transformation in coordinates
\bq\label{trans-var}
J_i=s_i+t_i,\qquad x_i=\varkappa(s_i-t_i),\qquad \varkappa\in \mathbb C\,,
\eq
looks like
\[J_1=\sqrt{u_2^2-u_1^2}\sin v_1 ,\qquad J_2=\sqrt{u_2^2-u_1^2}\cos v_1,
\qquad J_3=u_1,\]
and
\[\begin{array}{l}
x_1=\frac{\mathcal C_1 \sqrt{1-\frac{u_1^2}{u_2^2} } +u_1\sqrt{\mathcal C_2 -\varkappa^2 u_2^2-\frac{\mathcal C_1 ^2}{u_2^2}}\cos v_2}{u_2}\sin v_1
     +\sqrt{\mathcal C_2 -\varkappa^2 u_2^2-\frac{\mathcal C_1 ^2}{u_2^2}}\sin v_2\cos v_1\,,\\
     \\
x_2=\frac{\mathcal C_1 \sqrt{1-\frac{u_1^2}{u_2^2} } +u_1\sqrt{\mathcal C_2 -\varkappa^2 u_2^2-\frac{\mathcal C_1 ^2}{u_2^2}}\cos v_2 }{u_2}\cos v_1
     +\sqrt{\mathcal C_2 -\varkappa^2 u_2^2-\frac{\mathcal C_1 ^2}{u_2^2}}\sin v_2\sin v_1\,,\\
     \\
x_3=\mathcal C_1 \frac{u_1}{u_2^2}-\sqrt{1-\frac{u_1^2}{u_2^2}}\sqrt{\mathcal C_2 -\varkappa^2 u_2^2-\frac{\mathcal C_1 ^2}{u_2^2}}\cos v_2
\end{array}
\]
Here
\ben
\mathcal C_1&=&x_1J_1+x_2J_2+x_3J_3=\varkappa(C_1-C_2),\nn\\
\nn\\
\mathcal C_2&=&x_1^2+x_2^2+x_2^2+\varkappa^2(J_1^2+J_2^2+J_3^2)=2\varkappa^2(C_1+C_2).\nn
\en
In $(x,J)$-variables (\ref{trans-var}) canonical bivector $P_0$ (\ref{p0}) on $so^*(4)$ reads as
\[
 P_0=\left(\begin{array}{cc}\varkappa^2J_M&x_M\\x_M&
J_M\end{array}\right).
\]
At $\varkappa\to 0$ this bivector is reduced to the canonical bivector $P_0$ on $e^*(3)$, which is a dual space to the algebra $e(3)$ of Euclidean group $E(3)$. After contraction $\varkappa\to 0$  our variables $u,v$ (\ref{u1}-\ref{u2}) coincide with the Andoyer variables on $e^*(3)$ \cite{and23}.

The projection of the canonical bivector $P_0$  on the generic symplectic leaves of $M=so^*(4)$ or $M=e^*(3)$ in $(u,v)$-variables (\ref{u1}-\ref{u2}) looks like
\[
\widehat{P}_0=\left(
                  \begin{array}{cccc}
                    0 & 0 & 1 & 0 \\
                    0 & 0 & 0 & 1 \\
                    -1 & 0 & 0 & 0 \\
                    0 & -1 & 0 & 0
                  \end{array}
                \right)
\]
Using this analog of the Andoyer variables we can easy obtain  the projection $\widehat{P}_1$
of any bivector $P_1$ fulfilling (\ref{m-eq3}).

So, we can introduce symplectic form $\omega=\widehat{P}_0^{-1}$, recursion operator $N=\widehat{P}_1\widehat{P}_0^{-1}$ and  Darboux-Nijenhuis variables on the generic symplectic leaves of $M=so^*(4)$ for all the bivectors $P^{(m)}_1$ from the Section 3.

\subsection{First bivector}
Let us consider bivector $P_1^{(1)}$ (\ref{p1-1}) on  $M=so^*(4)$.
Using orthogonal transformations (\ref{ort-tr})
we can always put
 \[a=(0,0,a_3),\qquad b=(0,0,b_3).\]
In this case bivector $P_1^{(1)}$ gives rise to  the following Darboux-Nijenhuis coordinates
\bq\label{q-P1}
q_1=2\alpha a_3s_3,\qquad q_2=2\alpha b_3t_3\,,
\eq
and  momenta
\bq\label{p-P1}
\begin{array}{l}
p_1=\dfrac{1}{2\alpha\,a_3}\arctan\left(\dfrac{s_1}{s_2}\right)-\dfrac{\beta
a_3}{4\alpha^2\, b_3}\ln(a_3s_3-b_3t_3),\\
p_2=\dfrac{1}{2\alpha\,b_3}\arctan\left(\dfrac{t_1}{t_2}\right)+\dfrac{\beta\,a_3}{4\alpha^2\,b_3}\ln(a_3s_3-b_3t_3),
\end{array}
\eq
which satisfy relations (\ref{br-xy}). It means that projections of $P_0$ and $P_1$ on the generic symplectic leaves of $P_0$ are equal to
\bq\label{proj}
\widehat{P}_0=\left(
                  \begin{array}{cccc}
                    0 & 0 & 1 & 0 \\
                    0 & 0 & 0 & 1 \\
                    -1 & 0 & 0 & 0 \\
                    0 & -1 & 0 & 0
                  \end{array}
                \right),\qquad
\widehat{P}_1^{(1)}=\left(
                  \begin{array}{cccc}
                    0 & 0 & q_1 & 0 \\
                    0 & 0 & 0 & q_2 \\
                    -q_1 & 0 & 0 & 0 \\
                    0 & -q_2 & 0 & 0
                  \end{array}
                \right)
\eq
and, therefore, symplectic foliations of $P_0$ and $P_1$ coincide. On the other hand it means that generic symplectic leaves are regular semisimple  $\omega N$ manifolds \cite{fp02}.

\subsection{Second bivector}
Let us consider bivector $P_1^{(2)}$ (\ref{p1-2}) on  $M=so^*(4)$.
Using orthogonal transformations (\ref{ort-tr}) and scaling
(\ref{s-tr})
 we can always put
 \bq\label{bb}
a=(0,0,1),\qquad b=(b_1,ib_1,0)\,.
 \eq
In this case the Darboux-Nijenhuis coordinates are roots of
the minimal polynomial of the  recursion operator $N^{(2)}$
\bq\label{q-P2}
\mathcal  A(\lambda)=(\lambda-q_1)(\lambda-q_2)=
\lambda^2+(t_3-s_3)\lambda+(t_1+it_2)(s_1+is_2)b_1^2-s_3t_3
\eq
whereas the corresponding momenta are equal to
\bq\label{p-P2}
p_{1,2}=-i\ln \mathcal B(\lambda=q_{1,2}),
\eq
where
\[
\mathcal B(\lambda)=
\bigl(s_1-is_2+b_1^2(t_1+it_2)\bigr)\lambda+t_3(s_1-is_2)+b_1^2s_3(t_1+it_2)\,,
\]
As above these variables satisfy relations (\ref{br-xy}) and, therefore, symplectic foliations of $P_0$ and $P_1^{(2)}$ coincide.

According \cite{ts06a}, we can easy derive relations (\ref{br-xy}) from the following relations
\ben
\{\mathcal A(\lambda),\mathcal B(\mu)\}_k&=&\dfrac{i}{\lambda-\mu}\Bigl(\lambda^k \mathcal B(\lambda)\mathcal A(\mu)- \mu^k\mathcal A(\lambda)\mathcal B(\mu)\Bigr),\nn\\
\label{AB-rel}\\
\{\mathcal A(\lambda)\mathcal A(\mu)\}_k&=&\{\mathcal B(\lambda)\mathcal B(\mu)\}_k=0,\qquad k=0,1.\nn
\en

\subsection{Third bivector}
Let us consider bivector $P_1^{(3)}$ (\ref{p1-3}) on  $M=so^*(4)$.
Using orthogonal transformations (\ref{ort-tr}) and scaling
(\ref{s-tr})
 we can always put
 \bq\label{cc}
a=(0,0,1),\qquad b=(b_1,ib_1,0),\qquad c=(c_1,-ic_1,0)\,.
 \eq
In this case the Darboux-Nijenhuis coordinates are roots of the minimal polynomial
of the corresponding recursion operators $N^{(3)}$
\bq\label{q-P3}
\mathcal A(\lambda)=(\lambda-q_1)(\lambda-q_2)=\lambda^2-(s_3+t_3)\lambda+c_1b_1(t_1-it_2)(s_1+is_2)+s_3t_3\,.
\eq
The corresponding momenta are defined by
\bq \label{p-P3}
p_{1,2}=-i\ln \mathcal B(\lambda=q_{1,2}),
\eq
where
\[
\mathcal B(\lambda)=
\bigl((s_1-is_2)+c_1b_1(t_1-it_2)\bigr)\lambda+c_1b_1(t_1-it_2)s_3-(s_1-is_2)t_3\,.
\]
As above these variables satisfy relations (\ref{br-xy}) and, therefore, symplectic foliations of $P_0$ and $P_1^{(3)}$ coincide.

It is easy to see that  polynomial
$\mathcal A(\lambda)$ (\ref{q-P2})  and polynomial $\mathcal A(\lambda)$ (\ref{q-P3})
coincide after the following canonical transformation
\bq\label{k-trans}
t_2\to -t_2, \qquad t_3\to -t_3\,,\qquad \mbox{at}\quad \hat{b}_1^2=c_1b_1.
\eq
Here $\hat{b}_1$ is the entry of the vector $b$ (\ref{bb}), whereas $b_1,c1$ are the entry of the vectors $b$ and $c$ (\ref{cc})

So, the corresponding Darboux-Nijenhuis variables and, therefore, bivectors $P_1^{(2)}$ (\ref{p1-2}) and $P_1^{(3)}$ (\ref{p1-3})
are equivalent up to canonical transformations.

\subsection{Fourth bivector}
Let us consider bivector $P_1^{(4)}$ (\ref{p1-e}) on  $M=e^*(3)$.
The Darboux-Nijenhuis coordinates are roots of the following minimal
polynomial of the recursion operator $N$
\ben%
\mathcal{A}(\lambda) &=& \lambda^2 + \Bigl( \langle a, J \rangle - \langle a
\wedge b,x\rangle \alpha\Bigr)\lambda
+\frac{\alpha^2\langle a,b\rangle}{4}\Bigl(\langle a,x \rangle\langle b,x\rangle-\langle a\wedge xv,b\wedge x\rangle\Bigr)\nn\\
&+&\frac{\alpha}{2}\langle a,b\rangle \langle a\wedge x,J\rangle
-\frac{1}{2\alpha} \langle a\wedge x,J\rangle
+\frac{1}{4\alpha^2}\langle x,x\rangle
-\frac{1}{2}\langle a,x \rangle\langle b,x\rangle
\en
Linear canonical transformations of $e^*(3)$ consist of rotations
\bq
\label{rotE3}
x\to \alpha\, {\rm U}\, x\,,\qquad  J\to {\rm U}\,J\,,
\eq
where $\alpha$ is an arbitrary parameter and $U$ is an orthogonal  matrix,
and shifts
\bq
 x\to  x \,,\qquad  J\to  J+ {\rm S}\, x\
,\label{shiftE3}
\eq
where ${\rm S}$ is an arbitrary $3\!\times\!3$ skew-symmetric constant matrix \cite{kst03}.
Using these canonical  transformations of the vectors $x$ and $J$ we can always put
\[\mathcal A(\lambda)= \lambda^2+J_3\lambda+\frac{1}{2\alpha}(x_2J_1-x_1J_2)+\frac{1}{4\alpha^2}(x_1^2+x_2^2+x_3^2).
\]
However, in this case we could not  directly solve equations (\ref{br-xy}) with respect to the
functions $p_{1,2}(x,J)$.

Nevertheless, projections of $P_0$ and $P_1^{(4)}$ on the generic symplectic leaves of $P_0$
are given by (\ref{proj}) and, therefore, symplectic foliations of $P_0$ and $P_1^{(4)}$ coincide.

\subsection{Higher order Poisson  bivectors }

For all the considered above Poisson bivectors $P_1^{(m)}$ we can build the higher order Poisson structures by using the corresponding Darboux-Nijenhuis variables. Namely, if we restrict our bivectors $P_0$ and $P_1^{(m)}$ on their common symplectic leaves and postulate the following Poisson brackets between the Darboux-Nijenhuis variables
\[
\{q_i,q_j\}_k=\{p_i,p_j\}_k=0,\qquad \{q_i,p_j\}=q_i^k\delta_{ij},
\]
then after transformations $(q,p)\to (u,v)\to(s,t)$
one gets  $k+1$ order Poisson bivectors on $so^*(4)$.

As an example we present the cubic Poisson bivector
\bq\label{p2}
P_2^{(3)}=\left(
                        \begin{smallmatrix}
                           i\bigl(a^{-1}\langle a,s\rangle^2-\langle b,s\rangle\langle c, t\rangle\bigr) s_M,\quad & -\langle a,s+t\rangle\bigl[a^{-1}(a\wedge s)\otimes(a\wedge t)+(b\wedge s)\otimes(c\wedge t)\bigr]  \\
                        *  &
                         i\bigl(a^{-1}\langle a,t\rangle^2-\langle b,s\rangle\langle c, t\rangle\bigr) t_M
                        \end{smallmatrix}
                      \right),
\eq
associated with the  variables (\ref{q-P3})- (\ref{p-P3}) and, therefore, compatible with the quadratic tensor $P_1^{(3)}$ (\ref{p1-3}).

\section{Bi-integrable systems}
\setcounter{equation}{0}

\subsection{The Jacobi method}
In order to get bi-integrable systems on $M$ we can identify
Darboux-Nijenhuis variables with the separated variables and
substitute all the pairs of variables $q_j,p_j$ into the separated equations
\bq\label{sepeq-1}
\Phi_j(q_j,p_j,\alpha_1,\alpha_2)=0,\qquad j=1,2,
\eq
where $\Phi_j$ are functions on $p_j,q_j$ and two parameters $\alpha_{1,2}$ only.

According to the Jacobi theorem  if we solve  the separated equations (\ref{sepeq-1})
with respect to parameters $\alpha_{1,2}$ one gets a pair of
independent integrals of motion
\bq\label{m-int}
\alpha_{1,2}=H_{1,2}(p,q),
\eq
as functions on the phase space $M=so^*(4)$, which are in the bi-volution
\bq\label{bi-inv1}
\{H_1,H_2\}_0=\{H_1,H_2\}_1=0\,.
\eq
with respect to the Poisson brackets associated with
bivectors $P_0$ and $P_1$ (see Propositions 1-3 in \cite{ts07}).

For instance, substituting the Darboux-Nijenhuis variables $p_j,q_j$ (\ref{q-P1}-\ref{p-P1}) associated with the first bivector $P_1^{(1)}$ into the relations
\[
\Phi_1=2q_1^2+2V(p_1)-H_1-H_2, \qquad \Phi_2=2 q_2^2-H_1+H_2,
\]
where $V=e^{4i\alpha a_3p_1}+e^{-4i\alpha a_3p_1}$, then at $\beta=-i\alpha$, $b_3=a_3=1$ one gets
\[
H_{1,2}=(q_1^2\pm q_2^2)+V=
4\alpha^2(s_3^2\pm t_3^2)+(t_3-s_3)\dfrac{s_1+is_2}{s_1-is_2}+(t_3-s_3)^{-1}\dfrac{s_1-is_2}{s_1+
is_2}\,.
\]
Using other separated equations $\Phi_{1,2}=0$ we can get many other more complicated bi-integrable systems on $so^*(4)$.

\subsection{Quadratic integrals of motion}
In this Section we will substitute all the known pairs of integrals of motion $H_{1,2}$ on $so^*(4)$ into the equations (\ref{bi-inv1}) and try to found bi-integrable systems associated with one of the Poisson bivectors from  Section 3. So, in this Section we will forget about the Darboux-Nijenhuis coordinates and will start with integrals of motion listed in \cite{bm} and in \cite{sok06}, which have  different physical applications.

In order to describe these bi-integrable systems we prefer
to use $(x,J)$ coordinates (\ref{trans-var}) as in \cite{bm,sok06}.

\begin{prop}
If  $A=-i\varkappa a$, $|a|=1$ and  $B$ is an arbitrary vector, then at
\[b\wedge c+2ia=0\]
the following integrals of motion
\bq
\label{ts-sys}
\begin{array}{l}
H_1=\langle A,B\rangle |J|^2-2\langle A,J\rangle \langle B,J \rangle+\langle B, J\wedge x\rangle,\\ \\
H_2=\langle  B,J\rangle\Bigl(
2\langle  A,J\wedge x\rangle-\varkappa^2\langle  J,J\rangle+\langle  x,x\rangle \Bigr)
\end{array}
\eq
and
\bq\label{sok-sys}
\begin{array}{l}
\widetilde{H}_1=\langle  A,J\wedge(B\wedge J)\rangle+\langle  B,J\wedge x\rangle\\
\\
\widetilde{H}_2=\langle  J,B\rangle^2\Bigl(
 \langle  J\wedge A,J\wedge A\rangle
 +2\langle A,J\wedge x\rangle+\langle  x,x\rangle \Bigr)\,
\end{array}
\eq
are in bi-involution
\bq\label{bi-inv}
\{H_1,H_2\}_0=\{H_1,H_2\}_1
=\{\widetilde{H}_1,\widetilde{H}_2\}_0=\{\widetilde{H}_1,\widetilde{H}_2\}_0=0\,,
\eq
with respect to the Poisson brackets associated with $P_0$ (\ref{p0}) and $P_1$ (\ref{p1-3}).
\end{prop}
Integrable system with the cubic integrals of motion  $H_{2}$ (\ref{ts-sys}) has been proposed in  \cite{ts04}. For this system we know the Lax matrices and the separated variables \cite{ts04}.  The second integrable systems with fourth order integral of motion  $\widetilde{H}_{2}$ (\ref{sok-sys}) has been considered in \cite{sok04}.

According to \cite{fp02} the bi-involutivity of integrals of motion (\ref{bi-inv}) is equivalent to the existence of  non-degenerate control matrices $F$ and $\widetilde{F}$, such that
\bq
P_1dH_i=P_0\sum_{j=1}^2 F_{ij}\,dH_j,\qquad P_1d\widetilde{H}_i=P_0\sum_{j=1}^2 \widetilde{F}_{ij}\,d\widetilde{H}_j,\qquad i=1,2.
\eq
In our case they look like
\bq\label{F-mat}
F=\left(
    \begin{array}{cc}
      \langle a,J\rangle & \dfrac{i}{2\varkappa} \\ \\
      \dfrac{-iH_2}{2\varkappa\langle B,J\rangle} & 0
    \end{array}
  \right),\qquad
    \widetilde{F}=\left(
                  \begin{array}{cc}
                    \dfrac{\langle a,J\rangle}{2} & \dfrac{i}{4\varkappa\langle B,J\rangle} \\ \\
                    \dfrac{-i\widetilde{H}_2}{\varkappa\langle B,J\rangle} & \dfrac{\langle a,J\rangle}{2}
                  \end{array}
                \right)
\eq
Entries of  $F$ are polynomials, whereas one of the entries of $\widetilde{F}$ is  a rational function on  $so^*(4)$.

The eigenvalues of the matrices  $F$ and $\widetilde{F}$ coincide to each other. They are  roots of the common characteristic polynomial
\bq\label{pol-A}
\mathcal A(\lambda)=(\lambda-q_1)(\lambda-q_2)=\lambda^2-\langle a,J\rangle\lambda+\dfrac{\langle J,J\rangle}4-\dfrac{i\langle a,x\wedge J\rangle}{2\varkappa}-\dfrac{\langle x,x\rangle}{4\varkappa^2}\,.
\eq
Of course, this polynomial coincides with the minimal polynomial $\mathcal A^{(3)}$ of the recursion operator $N^{(3)}$ after suitable canonical transformation.

Using relations (\ref{AB-rel}) we can reconstruct the conjugated momenta $p_{1,2}$. Namely, if numerical vector $d$ satisfies conditions
$\langle a,d\rangle=\langle d,d\rangle=0$, the coordinates $q_{1,2}$ (\ref{pol-A}) and the momenta \bq\label{pol-B}
p_{1,2}=-i\ln \mathcal B(\lambda=q_{1,2}),\qquad \mathcal B(\lambda)=\{\langle d,J\rangle, \mathcal A(\lambda)\}_0
\eq
are the Darboux-Nijenhuis variables fulfilling equations (\ref{br-xy}).
The variables  $p_{1,2}$ (\ref{pol-B}) are defined up to canonical transformation
$p_i\to p_i+f_i(q_i)$, where $f_i$ are arbitrary functions on $q_i$ only.

\begin{prop}
 Coordinates $q_{1,2}$ (\ref{pol-A}) and momenta $p_{1,2}$ (\ref{pol-B}) are the separated variables for the integrable systems with integrals of motion $H_{1,2}$ (\ref{ts-sys}) and $\widetilde{H}_{1,2}$ (\ref{sok-sys}).
If
\[
b^*=c,\qquad \mbox{and}\qquad d=c,
\]
the corresponding separated equations are equal to
\ben
4\varkappa^2 \langle a,B\rangle\, q_k^3+q_kH_1-H_2&=&
2\varkappa^2\langle c,a\wedge B\rangle\,
(q_{k}^2-C_1)(q_{k}^2-C_2)\,q_k\,\e^{-ip_{k}}\nn\\
&+&2\varkappa^2\langle b,a\wedge B\rangle\,q_k\, \e^{ip_{k}}
\label{sepeq-par}
\en
and
\ben
2\varkappa\langle a,B\rangle\, q_{1,2}^2-\widetilde{H_1}\mp\sqrt{\widetilde{H_2}}&=&
\varkappa\langle c,a\wedge B\rangle\, (q_{1,2}^2-C_1)(q_{1,2}^2-C_2)\, \e^{-ip_{1,2}}
\nn\\
&+&\varkappa\langle b,a\wedge B\rangle\, \e^{ip_{1,2}}
\label{sepeq-car}
\en
\end{prop}
The separated equations (\ref{sepeq-par}) and (\ref{sepeq-car}) are related with the parabolic and cartesian St\"{a}ckel webs on a plane. The  St\"{a}ckel matrices
$S$ and $\widetilde{S}$ diagonalize the control matrices $F$ and $\widetilde{F}$:
\[F=S\left(\begin{array}{cc}
       q_1 & 0 \\
              0 & q_2
     \end{array}\right) S^{-1},\qquad
S=\left(
    \begin{array}{cc}
      1 & 1 \\
            2i\varkappa q_2 & 2i\varkappa q_1
    \end{array}
  \right)\,,
\]
and
\[\widetilde{F}=\widetilde{S}\left(\begin{array}{cc}
       q_1 & 0
       \\
       0 & q_2
     \end{array}\right) \widetilde{S}^{-1},\qquad
\widetilde{S}=\left(
    \begin{array}{cc}
      1 & 1
      \\
      2i\sqrt{\widetilde{H}_2} & -2i\sqrt{\widetilde{H}_2}
    \end{array}
  \right)\,.
\]
The right hand sides of the separated equations (\ref{sepeq-par}) and (\ref{sepeq-car}) are the generalized St\"{a}ckel potentials.

 For  special values of  $A$, $B$  and $\varkappa$ the Hamiltonians (\ref{ts-sys}) and (\ref{sok-sys})
are real functions \cite{ts04,sok04}. There is one integrable system with the complex Hamiltonians
\ben
\widehat{H}_1&=&\alpha J_2^2-\frac{\varkappa^2}{\alpha}J_1^2+x_2J_1-x_1J_2,\qquad \alpha=i\varkappa\,,\nn\\
\nn\\
\widehat{H}_2&=&\alpha(J_1x_2-J_2x_1)(\varkappa^2|J|^2-|x|^2)+\varkappa^2
\Bigl((J_1x_2-J_2x_1)^2+(J_3x_1-J_1x_3)^2\Bigr)\nn\\
&-&\alpha^2\Bigl((J_1x_2-J_2x_1)^2+(J_2x_3-J_3x_2)^2\Bigr)\nn
\en
which are in the bi-involution with respect to the same Poisson brackets at $a_1=a_2=0$ and $a_3=-1$. At the arbitrary value of $\alpha$ the Lax matrices and the separated variables for this system have been  constructed in  \cite{ts02,ts04}.

\subsection{Inhomogeneous integrals of motion}
There is only one nontrivial linear  Poisson bivector, which is compatible with the canonical bivector $P_0$ and the quadratic bivector $P_1^{(3)}$ (\ref{p1-3})
\[
P_1^{(0)}=\left(
                    \begin{array}{cc}
                      \alpha_1s_M & 0 \\
                      0 & \alpha_2 t_M
                    \end{array}
                  \right),\qquad \alpha_{1,2}\in\mathbb C.
\]
The linear combination
\bq\label{p1-g}
P_1^g=P_1^{(3)}+\left(
                    \begin{array}{cc}
                      \alpha_1s_M & 0 \\
                      0 & \alpha_2 t_M
                    \end{array}
                  \right)
\eq
is an inhomogeneous Poisson bivector on $so^*(4)$ compatible with $P_0$ and such that $P_1^gdC_{1,2}=0$.

 Let us consider the following inhomogeneous Hamiltonians  \cite{ts04,sok06}
\ben
H_1^g= H_1&+&\langle k,J\rangle +\alpha_{12}\langle B,x\rangle,\label{tsig-sys2}\\
\nn\\
H_2^g=H_2&+&2\alpha_{12}\langle B,J\rangle\langle,A,x\rangle-\langle k,A\rangle J^2-\langle k,J\wedge x\rangle\nn\\
&-&\varkappa^2\alpha_{12}^2\langle B,J\rangle-\alpha_{12}\langle k,x\rangle\,,\nn
\en
and
\ben
\widetilde{H}_1^g&=& \widetilde{H}_1+\langle \alpha_3 A+\alpha_4 A\wedge B,J\rangle +\alpha_{12}\langle B,x\rangle,\label{sok-sys2}\\
\nn\\
\widetilde{H}_2^g&=&\widetilde{H}_2+2 \langle B,J\rangle
\Bigl[
  \alpha_3 \bigl(\langle A,J\rangle ^2+\varkappa^2 |J|^2 -|x|^2 -2 \langle A,J\wedge x\rangle  \bigr)\nn\\
&+&\alpha_4 \bigl(\langle B\wedge A, J\wedge x\rangle  +\langle A,J\rangle  \langle A,B\wedge J\rangle \bigr)
+\alpha_{12} \langle B,J\rangle  \langle A,x\rangle
\Bigr]\nn\\
&+&\alpha_3^2\Bigl( 2 |x|^2 -\langle A,J\rangle ^2+2 \langle A,J\wedge x\rangle  \Bigr)
 +2 \alpha_4 \alpha_{12} \Bigl( \langle B,J\rangle  \langle A,B\wedge x\rangle     \Bigr)\nn\\
&+&2 \alpha_4^2 \Bigl( \langle B,B\rangle  \langle A,J\rangle ^2-\varkappa^2 \langle B,J\rangle ^2-2 \langle A,B\rangle  \langle A,J\rangle  \langle B,J\rangle  \Bigr)\nn\\
&-&4 \alpha_3 \alpha_{12} \langle B,J\rangle  \langle A,x\rangle
 -\alpha_{12}^2 \varkappa^2 \langle B,J\rangle ^2\nn\\
&+&2 \alpha_3 \alpha_{12} \Bigl( \langle A\wedge B,J\wedge x\rangle -\langle A,J\rangle  \langle A,B\wedge J\rangle   \Bigr)\nn\\
&+&2 \alpha_3 \alpha_{12} \Bigl(\alpha_3 \langle A,x\rangle +\alpha_4 \langle A,B\wedge x\rangle +\alpha_{12} \varkappa^2 \langle B,J\rangle \Bigr),\nn
\en
where $\alpha_{12}=i(\alpha_1-\alpha_2)$, $\alpha_{1,\ldots,4}$ are arbitrary parameters and $k$ is arbitrary numerical vector.

\begin{prop}
The inhomogeneous integrals of motion (\ref{tsig-sys2}) and (\ref{sok-sys2}) are in the bi-involution with respect to the canonical brackets $\{.,.\}_0$ and the second  Poisson bracket $\{.,.\}_1$ associated with the  bivector $P_1^g$ (\ref{p1-g})
\end{prop}
The separated variables for these inhomogeneous bi-integrable systems are different, but
they remain the eigenvalues of the corresponding control matrices
\[
F^g=F+\dfrac12\left(
                \begin{array}{cc}
                  \alpha_1+\alpha_2 & 0 \\ \\
                  \frac{\alpha_1-\alpha_2}{\varkappa}\langle k,x\rangle-\frac{i}{\varkappa}\langle k,J\wedge x\rangle-\langle k,B\rangle|J|^2 & \alpha_1+\alpha_2
                \end{array}
              \right)\,
\]
and
\[
\widetilde{F}^g=\left(
                  \begin{array}{cc}
                    \dfrac{\langle a,J\rangle + \alpha_1+\alpha_2}2 & \dfrac{i}{4\varkappa(\langle B,J\rangle-\alpha_3)} \\ \\
                    -\dfrac{i\widetilde{H}^g_2(\alpha_3=\alpha_4=0)\,(\langle B,J\rangle)-\alpha_3}{
                    \varkappa\langle B,J\rangle^2} & \dfrac{\langle a,J\rangle + \alpha_1+\alpha_2}2
                  \end{array}
                \right).
\]
For the inhomogeneous Hamiltonians
we have to add to the left hand side of the separated relations (\ref{sepeq-par}) and  (\ref{sepeq-car}) terms proportional  $q_k^2$ and  $q_k$ respectively. Moreover, in the right hand side  we have to substitute $((q_{k}-i\alpha_1)^2-C_1)((q_{k}-i\alpha_2)^2-C_2)$ instead of
$(q_{k}^2-C_1)(q_{k}^2-C_2)$.

\section{Conclusion}
\setcounter{equation}{0}
We classify  quadratic Poisson bivectors having the given canonical foliation on $M=so^*(4)$  or $M=e^*(3)$ as their symplectic leaf foliation. The corresponding Darboux-Nijenhuis variables are constructed. A pair of known integrable systems on $so^*(4)$ can be related with  one of such quadratic Poisson bivectors. We prove that the corresponding integrals of motion admit  separation of variables and the separated coordinates are eigenvalues of the control matrices.

Another approach to the construction of the quadratic and cubic Poisson bivectors $P_1$ having  common symplectic foliations with  $P_0$ has been proposed in \cite{ts07b,ts07c}. Among the corresponding  bi-integrable systems there are  generalized periodic Toda lattices, the  $XXX$ Heisenberg magnet,
the $XXX$ Heisenberg magnet with boundary conditions, the Kowalevski top on $so^*(4)$, the Goryachev-Chaplygin gyrostat on $e^*(3)$, the Kowalevski-Chaplygin-Goryachev gyrostat on $e^*(3)$ and some other well known integrable systems. Now we add to this list two other  integrable systems on $so^*(4)$.

The research was partially supported by
the RFBR grant 06-01-00140 and grant NSc 5403.2006.1.

\end{document}